\pdfoutput=1
\documentclass[10pt,oneside,pdftex,dvipsnames,a4paper]{article}

\usepackage{etex}
\reserveinserts{18}

\usepackage{a4wide}
\usepackage{setspace,xspace,makeidx,fancyvrb,relsize}
\usepackage{alltt}

\usepackage{amsmath,amsthm,amssymb}
\usepackage{parallel,amsfonts,accents}
\usepackage{amsxtra,amstext,latexsym,dsfont} % for \mathds{N}
\normalfont
\usepackage{bm}

\usepackage[comma]{natbib}
\usepackage[symbol]{footmisc}
\usepackage{perpage}
\MakePerPage[2]{footnote}

\usepackage[pdftex]{graphicx}
\usepackage[dvipsnames]{xcolor}

\usepackage{pgf,tikz}
\usetikzlibrary{snakes}
\usetikzlibrary{backgrounds}
\usetikzlibrary{arrows}
\usetikzlibrary{patterns}
\usetikzlibrary{fit}
\usetikzlibrary{calc}

\usepackage{lscape} 
\usepackage{shorttoc}

\usepackage{array,longtable,booktabs,float,rotating}
\usepackage{dcolumn,ctable}

\usepackage[labelfont={bf},ruled,labelsep=period,small]{caption}

\usepackage[flushleft]{threeparttable}
\usepackage[linkcolor=black,colorlinks=true,citecolor=black,%ForestGreen,%pagebackref,%
bookmarks=true,bookmarksopen=true,bookmarksnumbered=true,bookmarksopenlevel=0,%
hyperindex=true,pdfpagemode=UseOutlines,linktocpage=true,pdfpagelayout=TwoColumnLeft,
pdftitle=Biostatistics,pdfstartview=FitH,pdffitwindow=true,urlcolor=blue]{hyperref}
\usepackage{ragged2e}

\usepackage{multirow,multicol}

\usepackage{fancyhdr,extramarks}

 \makeatletter%--------------------------
 \newcommand\sub{\@startsection%
     {subsubsection}{5}{0mm}{-1\baselineskip}{.01\baselineskip}%
     {\normalfont\itshape}}
 \renewcommand\subsubsection{\@startsection%
     {subsubsection}{3}{0mm}{-1\baselineskip}{.01\baselineskip}%
     {\normalfont\itshape}}
 \makeatother%--------------------------

\makeatletter
        \newcommand\Appendix[2][?]{%
            \refstepcounter{section}%
            \addcontentsline{toc}{appendix}%
                {\protect\numberline{\appendixname~\thesection}#1}%
            {\raggedleft\bfseries \appendixname\
                \thesection\par \centering#2\par}%
                \sectionmark{#1}%
                \@afterheading
                \addvspace{\baselineskip}}
        \newcommand\sAppendix[1]{%
            \raggedleft\bfseries\appendixname\par
            \@afterheading\addvspace{\baselineskip}}
    \makeatother

\normalsize \makeindex

\newcolumntype{A}{>{\centering}p{100pt}}
\newlength\savedwidth

\def\coldot{.}%
{\catcode`\.=\active%
    \gdef.{$\egroup\setbox2=\hbox to \dimen0 \bgroup$\coldot}}
\def\rightdots#1{%
    \setbox0=\hbox{$1$}\dimen0=#1\wd0%
    \setbox0=\hbox{$\coldot$}\advance\dimen0 \wd0%
    \setbox2=\hbox to \dimen0 {}%
    \setbox0=\hbox\bgroup\mathcode`\.="8000 $}
\def\endrightdots{$\hfil\egroup\box0\box2}

\newcolumntype{d}[1]{D{.}{.}{#1}}
\newcolumntype{A}{>{\centering}p{100pt}}
\newcolumntype{.}{D{.}{.}{-1}}
\newcolumntype{P}[2]{>{#1\raggedright\arraybackslash}p{#2}}

\DeclareFontFamily{U}{euc}{}% I chose euc because the chart is called Euler cursive
\DeclareFontShape{U}{euc}{m}{n}{<-6>eurm5<6-8>eurm7<8->eurm10}{}%
\theoremstyle{plain}      
\theoremstyle{plain}      
\theoremstyle{plain}      
\theoremstyle{definition} 
\theoremstyle{definition} 
\theoremstyle{definition} 
\theoremstyle{plain} 
\theoremstyle{definition} 
\theoremstyle{plain} 
\theoremstyle{definition} 

\newcounter{nctr}
\usepackage[rightbars,color]{changebar}
\cbcolor{blue}
\VerbatimFootnotes
\usepackage{url}
\newcommand\tb{\textbf}
\newcommand\ti{\textit}

\usepackage{fancyhdr}
\begin{document}
\sloppy
% Spacing.
%\onehalfspacing
%\doublespacing
%\setstretch{3} 
\setcounter{secnumdepth}{-2}

%-----------------------------------------------
% Title Page
\title{Topological Randomness and Number of Edges
  Predict Modular Structure in Functional Brain Networks.}
\author{Ginestet C.E., O'Muircheartaigh, J., O'Daly, O.G. and Simmons, A.}
\date{}
\maketitle
%-----------------------------------------------
\setcounter{secnumdepth}{-2}

%%% Directions:
% Deadline: 18 July. 
% Main text should be about 500 words. No more, although it is
% stipulated as 250 words in the information for authors.
% In addition, we should produce about one figure and one table.
% This should not include more than 5 references.
% Overall, this cannot take more than 2 pages.

%%%%%%%%%%%%%%%%%%%%%%%%%%%%%%%%%%
%%\sub{Introduction}
In a recent paper, \citet{Bassett2011} have analyzed the static and dynamic organization 
of functional brain networks in humans. We here focus on the
first claim made in this paper, which states that the static modular
structure of such networks is nested with respect to time. 
\citet{Bassett2011} argue that this graded structure underlines a
``multiscale modular structure''.

In this letter, however, we show that such a relationship is substantially
mediated by an increase in the random variation of the correlation
coefficients computed at different time scales. In page 8 of their
Supplementary Information, \citet{Bassett2011} report that the size of
the mean correlation diminishes with the
size of the time window. Such a decrease in overall correlation 
will generally have two effects: (i) networks'
topologies will become increasingly more random and (ii) the number of
significant edges will decrease. In this letter, we use synthetic
data sets to show that these two phenomena are likely to be associated
with a higher number of modules, thereby potentially explaining the apparent
multiscale modular structure described by \citet{Bassett2011}. Our
simulations are based on the unweighted unsigned version of the
modularity algorithm of \citet{Clauset2004}, but may be extrapolated
to weighted signed adjacency matrices.
%%%%%%%%%%%%%%%%%%%%%%%%%%%%%%%%%%
\begin{figure}[t]  
  {\Large\tb{A}}\hspace{6.3cm}{\Large\tb{B}}\\ 
  \includegraphics{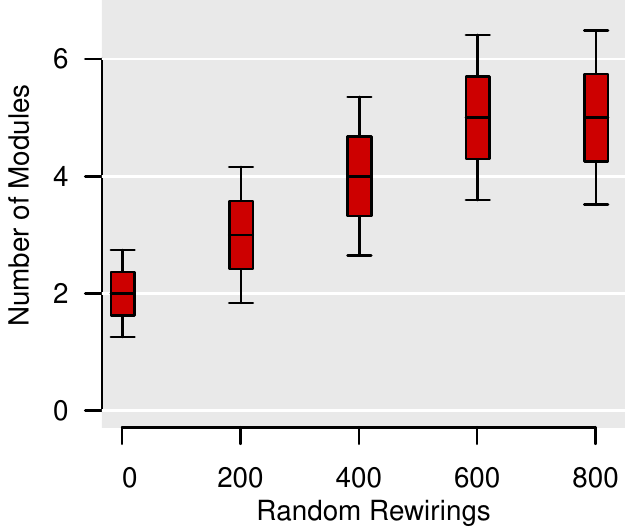}
  \hspace{.3cm}              
  \includegraphics{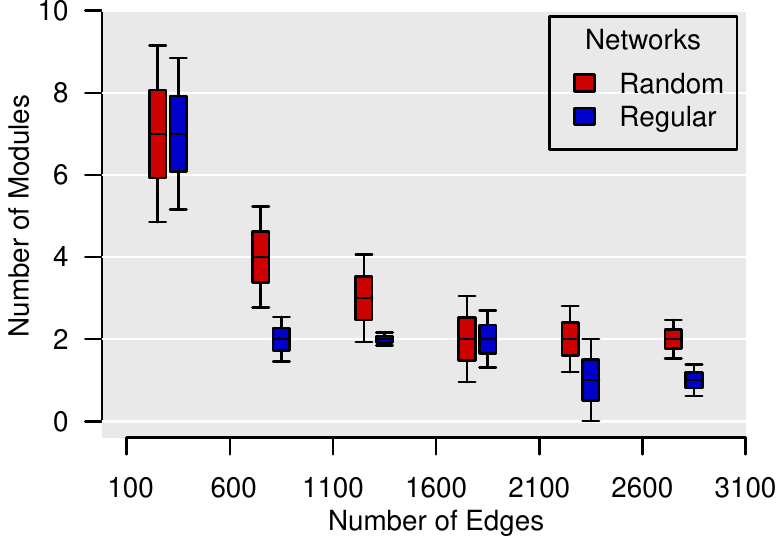} 
  \\
  {\Large\tb{C}}\vspace{-.5cm}
  \begin{center}
    \includegraphics[width=15cm]{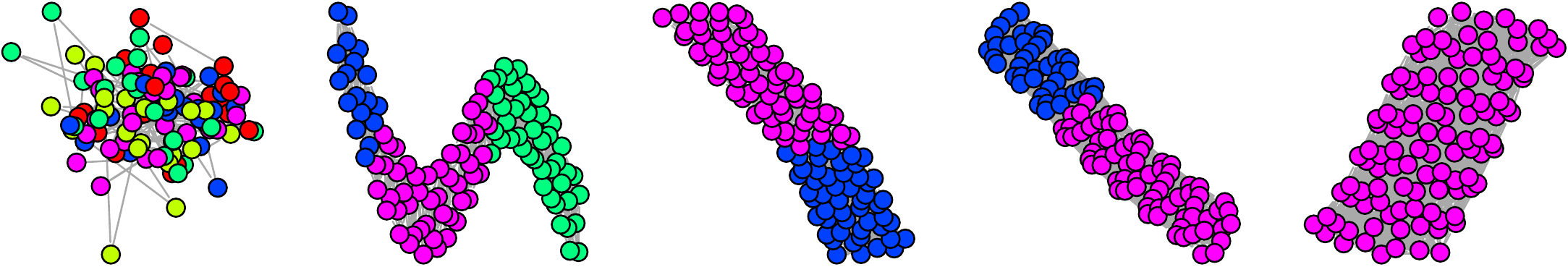}
  {\footnotesize
  \tb{$N_{E}=100$}\hspace{1.5cm}
  \tb{$N_{E}=600$}\hspace{1.5cm}
  \tb{$N_{E}=1100$}\hspace{1.5cm}
  \tb{$N_{E}=1600$}\hspace{1.5cm}
  \tb{$N_{E}=2100$}}
  \end{center}
  \vspace{.5cm}
  {\Large\tb{D}}
  \vspace{-.5cm}
  \begin{center}
    \includegraphics[width=15cm]{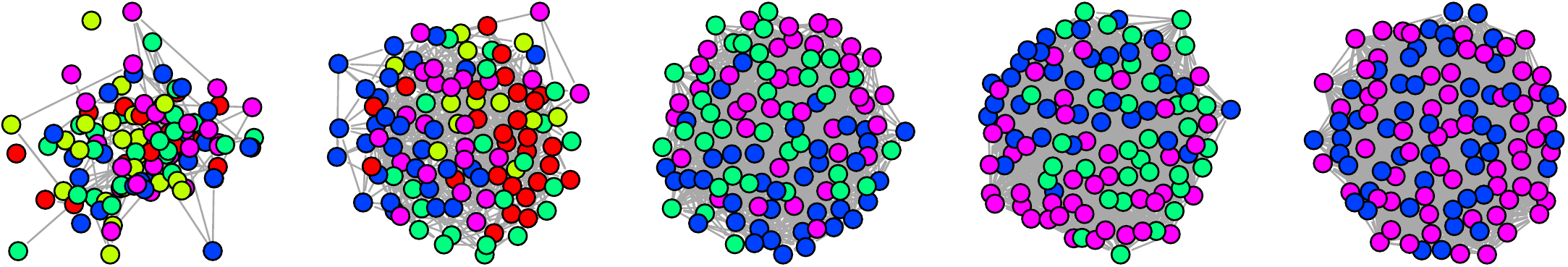}
  {\footnotesize
  \tb{$N_{E}=100$}\hspace{1.5cm}
  \tb{$N_{E}=600$}\hspace{1.5cm}
  \tb{$N_{E}=1100$}\hspace{1.5cm}
  \tb{$N_{E}=1600$}\hspace{1.5cm}
  \tb{$N_{E}=2100$}}
  \end{center}
  \caption{\tb{Topological randomness and number of edges predict number of
      modules.} (A) Relationship between the number of random rewirings
    of a regular lattice and the number of modules in such a
    network. Here, the number of edges is kept constant throughout all
    rewirings. (B) Relationship between the number of edges in a
    network and its number of modules for both regular (i.e. lattice) and random
    graphs. This shows that the number of modules tends to decrease as
    more edges are added to both types of networks. (C-D)
    Modular structures of regular (C) and random (D) networks for
    different number of edges, $N_{E}$. These networks are represented using the algorithm of 
    \citet{Kamada1989} with different colors representing different
    modules. In all simulations, the number of vertices is
    $N_{V}=112$.\label{fig:nets}}
\end{figure}
%%%%%%%%%%%%%%%%%%%%%%%%%%%%%%%%%%
In panel (A) of figure \ref{fig:nets}, we have generated 1,000 unweighted
lattices based on 112 vertices as in \citet{Bassett2011}. By randomly
rewiring the edges of these lattices, we show that 
the number of modules in these networks tends to increase with the
level of topological randomness in these graphs.
For panels (B) to (D), we have generated two sets
of unweighted networks, characterized by a random and a regular
topology, respectively, with different number of edges. 
These simulations were repeated 1,000 times for each type of graph for
each number of edges. For  both types of networks, the number of
modules in these graphs tended to decrease as new edges were added. 
Collectively, although these data simulations do not entirely rule out
the possibility of a temporally nested modular structure in the human brain,
they nonetheless cast doubts on the possibility of detecting such a temporal
organization by reducing the size of the sampling window. 

Note that our discussion, in this letter, has solely been concerned
with \ti{static} brain
networks. Thus, this critique does not call into question the two
other main conclusions of the paper published by \citet{Bassett2011},
which pertain to the \ti{dynamic} properties of these brain networks.

%%%%%%%%%%%%%%%%%%%%%%%%%%%%%%%%%%
%%%%%%%%%%%%%%%%%%%%%%%%%%%%%%%%%%
%%%%%%%%%%%%%%%%%%%%%%%%%%%%%%%%%%
\footnotesize
\sub{Acknowledgements}
CEG and AS are supported by the UK National Institute
for Health Research (NIHR) Biomedical Research Centre for Mental
Health (BRC-MH) at the South London and Maudsley NHS Foundation Trust
and King's College London and by the Guy's and St Thomas' Charitable
Foundation as well as the South London and Maudsley Trustees. 
% We thank \citet{Bassett2011} for making their data set available. 

% references --------------------------------------------------
\footnotesize
\singlespacing
\addcontentsline{toc}{section}{References}
\bibliography{/home/cgineste/ref/bibtex/Statistics,%
             /home/cgineste/ref/bibtex/Neuroscience}
\bibliographystyle{/home/cgineste/ref/style/oupced3}

% Index -------------------------------------------------------

\addcontentsline{toc}{section}{Index}
%\printindex

\end{document}